\documentclass{article}
\title{Cartan's Torsion: Necessity and Observational Evidence}
\author{Rainer W. K\"uhne \\ 
Lechstr. 63, 38120 Braunschweig, Germany \\ kuehne70@gmx.de}
\begin{document}
\maketitle

%\topmargin=0mm
%\oddsidemargin=0pt
%\textwidth=160.0mm
%\textheight=170.0mm
%\begin{document}

%\begin{center}
%{\Large {\bf Cartan's Torsion: Necessity and Observational Evidence}}

%\vspace{0.5cm}

%Rainer W. K\"uhne \\
%{\it Lechstr. 63, 38120 Braunschweig, Germany} \\
%kuehne70@gmx.de

%\end{center}

%\vspace{0.5cm}

%\noindent
This article starts with the mathematical definition, concrete description, 
and physical meaning of Cartan's torsion. I proceed with the argumentation
that torsion is required for the description of intrinsic spin. Moreover I
argue that the duality between curvature and torsion is analogous to the
duality between electricity and magnetism. I conclude this article by
pointing out that the aligned rotation axes of the galaxies of the
Perseus-Pisces supercluster may be interpreted as a topological defect
generated by torsion.

\section{What is Cartan's Torsion?}

When a four-vector $C^k$ is parallely displaced from the four-position
$x^k$ to $x^k + dx^k$, then it changes according to the prescription,
\begin{equation}
dC^k= - \Gamma^{k}_{ij} (x) C^j dx^{i} .
\end{equation}
This is the definition for the position-dependent affine connection
$\Gamma^{k}_{ij}$. According to general relativity \cite{1}, it has only
a symmetric part,
\begin{equation}
\{ \}^{k}_{ij} = \frac{1}{2} ( \Gamma^{k}_{ij} + \Gamma^{k}_{ji} ),
\end{equation}
which is called "Christoffel symbol." The anti-symmetric part of the
affine connection is called "Cartan's torsion" \cite{2},
\begin{equation}
T^{k}_{ij} = \frac{1}{2} ( \Gamma^{k}_{ij} - \Gamma^{k}_{ji} ).
\end{equation}
According to general relativity, the torsion tensor is zero. The
introduction
of a nonzero torsion tensor means therefore an extension of general
relativity.

Quite remarkably, the torsion tensor transforms as a tensor under local
Lorentz transformations, whereas the Christoffel symbol does not.

The torsion tensor can be viewed as the translational field strength.
It represents a closure failure of infinitesimal displacements. In
spacetimes
which include torsion, infinitesimal parallelograms do not close.

We know from Einstein's general relativity \cite{1} that gravitational mass
is connected with curvature via
\begin{equation}
G^{ij}= \kappa\Sigma^{ij},
\end{equation}
where
\begin{equation}
G^{ij}= R^{ij} - \frac{1}{2} g^{ij}R^{k}_{k}
\end{equation}
is the Einstein tensor, $\Sigma^{ij}$ is the stress-energy (energy-momentum)
tensor, $R^{ij}$ is the Ricci tensor, $g^{ij}$ is the metric tensor,
$R^{k}_{k}$ is the Ricci scalar, and $\kappa = -8\pi G/c^4$ is the
Einstein constant.

Analogously, intrinsic spin is connected with Cartan's torsion via
\begin{equation}
T^{ijk} = \kappa\tau^{ijk},
\end{equation}
where $\tau^{ijk}$ is the spin tensor \cite{3}. The equations (4) and (6)
show the duality between mass and spin and between curvature and torsion,
respectively.

Directly from the definition of the affine connection, Eq. (1), one 
obtains the differential equation of autoparallel curves,
\begin{equation}
\frac{d^2x^k}{ds^2} + \Gamma^{k}_{ij} \frac{dx^{i}}{ds}
\frac{dx^{j}}{ds} =0,
\end{equation}
where the infinitesimal interval $ds$ between $x^k$ and $x^k +dx^k$ is
given by
\begin{equation}
 ds^2 = g_{ij}(x) dx^{i}dx^{j}.
\end{equation}
Quite remarkably, only the symmetric part of the metric tensor contributes
to the square of the infinitesimal interval.

Readers who would like to learn more about the formalism of torsion
are invited to read the excellent reviews, Ref. \cite{4}.

\section{Why Do We Need Torsion?}

The energy-momentum tensor $\Sigma^{ij}$ of a Dirac field $\Psi$
(spin 1/2 field \cite{5}) is anti-symmetric \cite{6},
\begin{equation}
\Sigma^{ij} = - \frac{\hbar c}{2} [ ( \nabla^{i} \bar\Psi )
\gamma^{j} \Psi - \bar\Psi \gamma^{j} \nabla^{i} \Psi ] ,
\end{equation}
where
\begin{equation}
\nabla_{i}= \partial_{i} +ieA_{i}
\end{equation}
is the covariant derivative. By contrast, the energy-momentum tensor
of general relativity \cite{1} is symmetric. In order to couple a
spinor field (Dirac field) to a gravitational field, one has to use an
energy-momentum tensor which includes anti-symmetric parts. Therefore
general relativity has to be generalized by the introduction of
Cartan's torsion \cite{3}.

I have shown that the duality between mass and spin is analogous to the
duality between electric charge and magnetic charge \cite{7}. The
electric-magnetic duality is,
\begin{eqnarray}
J^{i} & = & \partial_{j} F^{ji} \\
j^{i} & = & \partial_{j} f^{ji} ,
\end{eqnarray}
where $J^{i}$ is the electric four-current, $j^{i}$ is the magnetic
four-current, and the field strength tensors are given by,
\begin{eqnarray}
F^{ji} & = & \partial^{j}A^{i} - \partial^{i}A^{j} \\
f^{ji} & = & \partial^{j}a^{i} - \partial^{i}a^{j} ,
\end{eqnarray}
where $A^{j}$ is the electric four-potential which corresponds to
Einstein's electric photon \cite{8}, and $a^{j}$ is the magnetic
four-potential which corresponds to Salam's magnetic photon \cite{9}.

Comparison of Eqs. (11) and (12) with Eqs. (4) and (6) demonstrates the
analogy between the electric-magnetic duality and the mass-spin duality.

The electric-magnetic duality is required to explain the quantization of
electric charge \cite{10}. Quantum field theoretical models which include
the magnetic photon can be found in Ref. \cite{11}. I argued \cite{12}
that magnetic photon radiation may have already been observed by
August Kundt in 1885 \cite{13}.

It is probably interesting to note that a Maxwell field $A^{i}$ which is
coupled to a gravitational field which includes both Cartan's torsion
\cite{2} and Weyl's non-metricity \cite{14}, requires the appearence of the
second four-potential $a^{i}$ \cite{15}.

Furthermore, Cartan's torsion tensor can be built from two 
independent vector fields which appear to obey the modified Maxwell 
equations of the two-photon theory \cite{15a}.

\section{Is There Observational \\ Evidence for Torsion?}

The rotation axes of the galaxies of the Perseus-Pisces supercluster are
aligned. This alignment exists over a distance of at least 40 Mpc
(130 million light years) \cite{16}. Such a large alignment cannot be
explained within the framework of conventional models of
galaxy-formation. Therefore I suggested \cite{17} that this alignment
is either a topological defect (torsion wall \cite{18}) or a remnant of the
original aligned distribution of galactic rotation axes generated by a
rotating universe \cite{19}. My interpretation of this structure as a
torsion
wall \cite{20} or as an effect of a rotating universe \cite{21} is
now generally accepted.

\end{document}